\documentclass[review]{elsarticle}

\usepackage{hyperref}
\usepackage[utf8]{inputenc}
\usepackage{subcaption}
\usepackage{float}
\usepackage{graphicx}
\usepackage[english]{babel}
\usepackage{amsmath}
\usepackage{amsfonts}
\usepackage{amssymb}
\usepackage{graphicx}
\usepackage{xcolor}
\journal{Neurocomputing}

\begin{document}

\begin{frontmatter}

\title{Data-driven simulation of pedestrian collision avoidance with a nonparametric neural network}

\author[mymainaddress]{Rafael F. Martin \corref{mycorrespondingauthor}}
\author[mysecondaryaddress]{Daniel R. Parisi}
\cortext[mycorrespondingauthor]{Corresponding author:ramartin@itba.edu.ar}

\address[mymainaddress]{Instituto Tecnol\'{o}gico de Buenos Aires. Lavarden 389, (C1437FBG) C. A. de Buenos Aires, Argentina.}
\address[mysecondaryaddress]{Instituto Tecnol\'{o}gico de Buenos Aires. CONICET. Lavarden 389, (C1437FBG) C. A. de Buenos Aires, Argentina.}

\begin{abstract}
Data-driven simulation of pedestrian dynamics is an incipient and promising approach for building reliable microscopic pedestrian models. We propose a methodology based on generalized regression neural networks, which does not have to deal with a huge number of free parameters as in the case of multilayer neural networks. Although the method is general, we focus on the one pedestrian - one obstacle problem. Experimental data were collected in a motion capture laboratory providing high-precision trajectories. The proposed model allows us to simulate the trajectory of a pedestrian avoiding an obstacle from any direction.
\end{abstract}


\end{frontmatter}

\section{Introduction}

We recently proposed a general framework of pedestrian simulation \cite{MartinPED20018} in which the surroundings of a virtual pedestrian, i.e., obstacles and other noncontacting particles, can only influence its trajectory by modifying its desired velocity. 

The basic assumption is that the avoidance behavior can be exerted only by the self-propelled mechanism of the particle itself (usually modeled by the desired velocity).

This framework is independent of the type of low-level model being force-based, rule-based or other. For example, it could be implemented on the Social Force Model \cite{helbing1995social, helbing2000simulating}, by replacing the social force, with a variable desired velocity that takes into account the possible future collisions \cite{qian2015new}. 

Of course, this framework can also be implemented on a first-order model, in which the position ($\textbf{r}$) of any particle can be updated by [ $\textbf{r}(t+\Delta t) = \textbf{r} (t) + \textbf{v}(t) \Delta t$ ] by dynamically adjusting the desired velocity [$\textbf{v}(t)$].

Under this approach, the problem lies in postulating the heuristics required for computing the variable desired velocity depending on the environment. As in traditional pedestrian theoretical models, any arbitrary heuristic can be proposed (for example, \cite{moussaid2011simple}, \cite{seitz2016cognitive}) and then the free parameters could be tuned in order to obtain simulated trajectories that approach experimental micro or macroscopic data. 

Instead of this traditional methodology, we can directly use the experimental data so as to compute the desired velocity at each time step. More precisely, we postulate that a minimum set of real trajectories exist, which could have the complete information for providing a desired velocity to the simulated agent, considering the state of the agent and its surroundings in the simulated and experimental environment. This concept of using experimental data in a simulated environment, is known as data-driven simulation.

Alternatively, the problem of simulating pedestrian dynamics can be seen in another dimension. The purpose of any model is to map current positions of particles [$ \textbf{r} (t) $, sometimes called `state' or `input'] into the positions at the next time step [$\textbf{r}(t+\Delta t)$, or `action' or `output']. Again, this mapping can be achieved by the large number of traditional pedestrian models \cite{schadschneider2009evacuation} and also by using available experimental data.

In this sense, data-driven simulations are becoming a new practice that has the benefit of avoiding proposal of explicit models along with their parameters, which can be related or not to reality. Instead, the experimental data can be considered directly for mapping the past positions into the future ones. No \textit{a priori} model assumption or guess needs to be made when simulating the pedestrian system. All the necessary information would be provided by the data from real scenarios.

Previous research papers using this approach have exploited the data directly \cite{kim2016interactive, zhao2015clustering, kouskoulis2018pedestrian} or through artificial neural networks (NN) \cite{song2018data, wei2018learning, alahi2016social, tordeux2018prediction, ma2016artificial}. This computation paradigm is a natural choice, because the NN can be "trained" with the experimental data and then it could be applied in the simulation when mapping old positions [$\textbf{r}(t)$] into the future ones [$\textbf{r}(t+\Delta t)$].

In general, these previous papers consider particular data sets corresponding to specific configurations (geometry and pedestrian flow) and then use these data for simulating the same system configuration. How the data-driven methodology could be generalized for simulating arbitrary (previously unseen) geometries and flow remains an open question.

In the case of using neural networks, all the previous papers implement Multilayer Perceptron \cite{haykin1994neural} (MLP, also known as Feed-Forward or Back-Propagation Neural Networks). These kinds of networks are very popular but their architecture presents an arbitrary number of hidden layers, each one with an arbitrary number of neurons. This leads to an also arbitrary number of free parameters (known as `weights') that should be determined via the training process using `input/output' pairs (called `patterns') from experimental data of the real system. 
For this reason, the number of patterns has to be much greater than the number of free parameters in order to find a reliable set of weights. In any case, by using an MLP the data are interpolated with a model having a huge number of parameters, which is, of course, an undesirable property for any model.

In this work, we propose a data-driven approach using a nonparametric universal interpolator: the generalized regression neural network (GRNN) \cite{specht1991general} (Sec. \ref{GRNN}). The GRNN needs to have access to the data examples when predicting a new output. However, because it has only one degree of freedom, the number of (input/output) patterns can be relatively low. And this brings us to the second novelty. We postulate that a complete set of (input/output) examples, extracted from a limited number of experimental trajectories, could be sufficient for simulating and reproducing any arbitrary pedestrian dynamic configuration. As a starting point, here we present this methodology in the case of one pedestrian avoiding a fixed obstacle. As the methodology is general, it will be implemented in more general scenarios of pedestrian dynamics in future work.

\section{The data-driven model}

In this section, the proposed data-driven model is explained in detail.

\subsection{General framework}\label{framework}
\begin{figure}[h]
\centering
\includegraphics[scale=0.5]{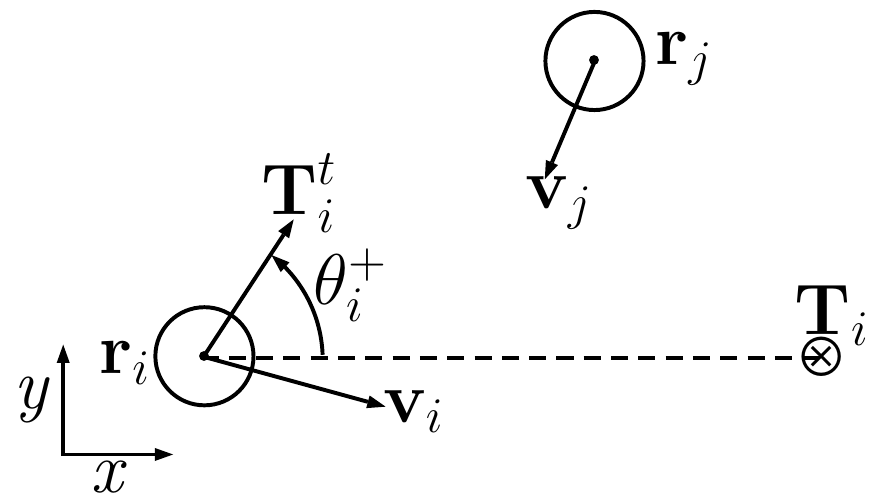}
\caption{Basic quantities of the general framework defining the environment of particle $i$ and the placement of the temporary local goal.} 
\label{Fig1}
\end{figure}

Our general framework \cite{MartinPED20018} postulates that a particle $i$, with position $\textbf{r}_i$, has a temporary and short-range goal $\textbf{T}_{i}^t(t)$ that is dynamically placed depending on the environment. $\textbf{T}_{i}^t$ will produce a detour in the trajectory in order to avoid any collision.  
	The environment is defined by the fixed long-distance goal ($\textbf{T}_i$), the positions ($\textbf{r}_j$) and the relative velocities ($\textbf{v}_{ij}$) of the nearest neighbors and obstacles. A graphical representation of these variables is presented in  Fig.\ref{Fig1}. 
	
We denote the general function that receives these relevant variables and returns the temporary and short-range goal as $\textbf{H}$:
	
\begin{equation}
\label{Heuristic}
\textbf{T}_{i}^t(t)= \textbf{H}(\textbf{r}_i(t),\textbf{r}_j(t),\textbf{v}_{ij}(t),\textbf{T}_i)
\end{equation}
\vspace{0 pt}

	The vector $\textbf{T}_{i}^t$ determines the avoidance direction, but it also has a magnitude that allows us to adjust the speed of the agent.
	The function $\textbf{H}$ is completely general and, of course, it can take any form. Again, we remark that this formulation does not depend on the type of low-level operational model. Thus, we choose a first-order model for describing the evolution of the particle, because we are not considering any forces, and besides, it presents a higher computation speed than a second-order model.
	
	The simulated particle has position	$\textbf{r}^s(t)$ and velocity $\textbf{v}^s(t)$ at time $t$ with a fixed long-distance goal $\textbf{T}$.

\begin{equation}
\label{first order equation}
	\textbf{r}^s(t+\Delta t)=\textbf{r}^s(t)+\textbf{v}^s(t)\Delta t,
\end{equation}
	 
\vspace{0 pt}	
where, we can directly identify the dynamic target with the desired velocity $\textbf{v}^s(t)=\textbf{T}^t(t)$.

\subsection{A minimum set of experimental trajectories}
\label{sec2p2}

Because this is a data-driven model, the experimental data are the first ingredient needed for obtaining the velocity ($\textbf{v}^s$) in eq. \ref{first order equation}.

As a case study of the proposed method, we will focus on a simple configuration, considering one pedestrian and one fixed obstacle.

\begin{figure}[h]
  \begin{subfigure}[t]{0.5\textwidth}
  \centering
    \includegraphics[scale=0.25]{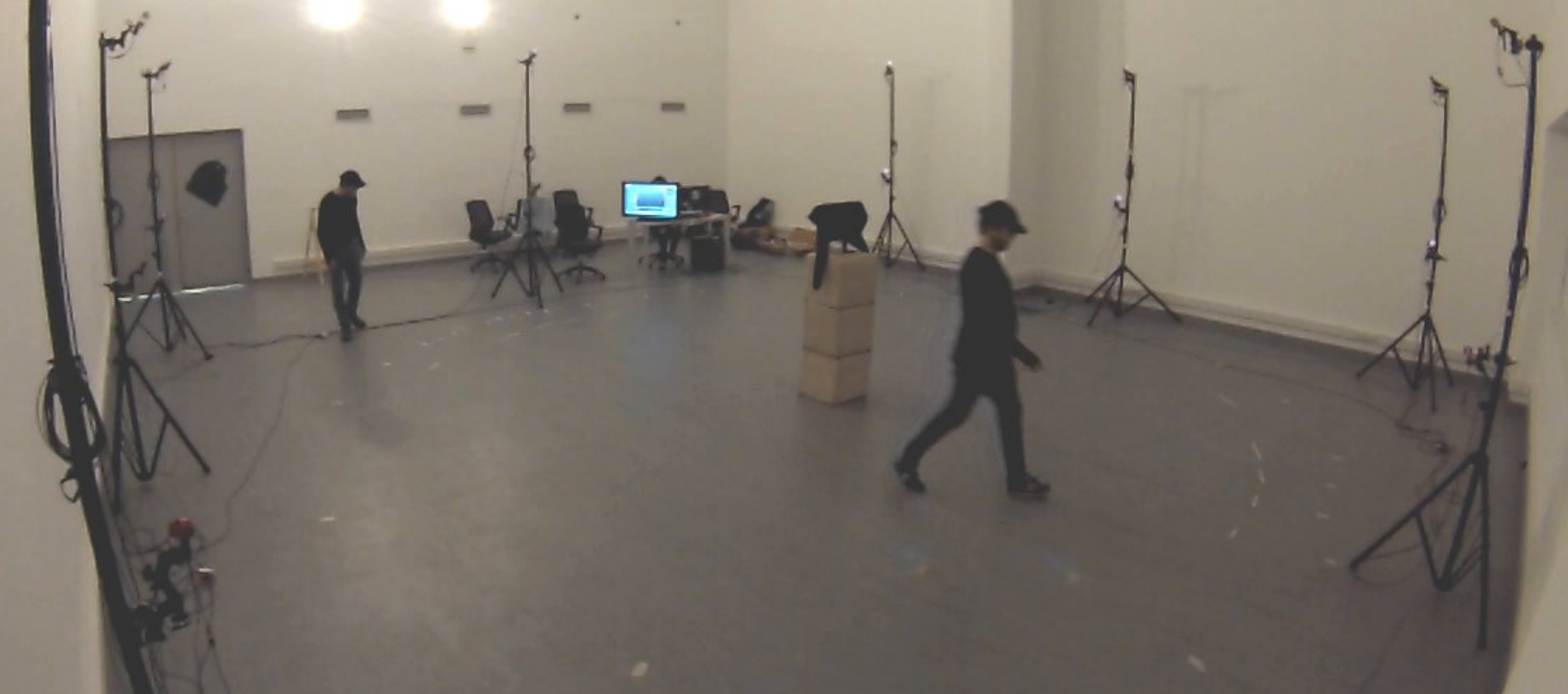}
	\caption{\centering{} }
  \end{subfigure}
  \begin{subfigure}[t]{0.6\textwidth}
  \centering
    \includegraphics[scale=0.45]{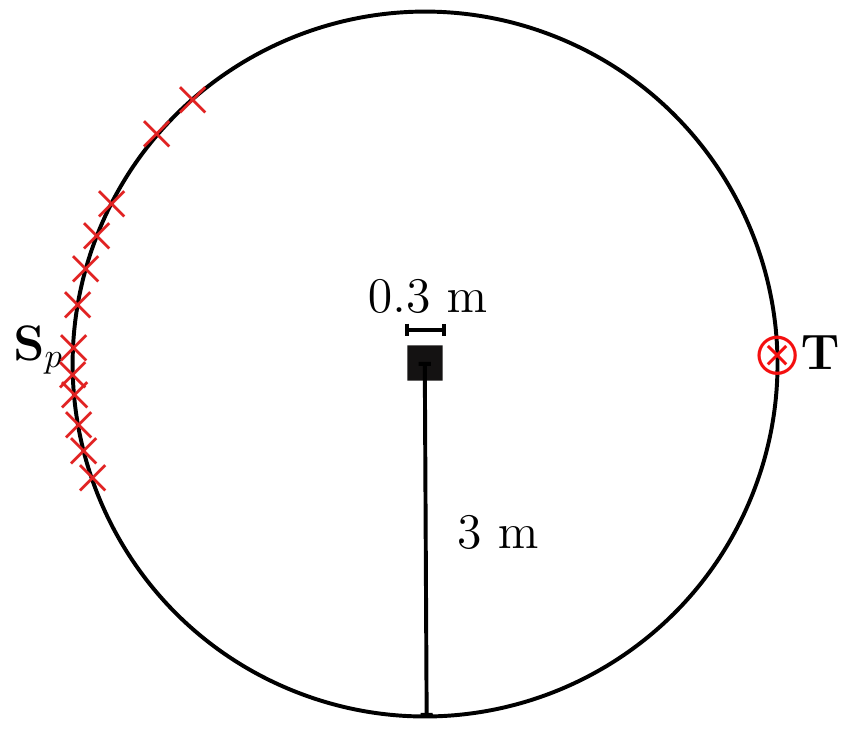}
	\caption{\centering{}}
  \end{subfigure}
   
  \caption{Experimental setting. (a) Snapshot of the experiment. (b) Schematic representation of starting points ($\textbf{S}_p$) and final target  ($\textbf{T}$) for the recorded trajectories.}

 \label{Fig2} 
 \end{figure}

The experimental setup consists of a 6 m circumference where we locate the starting points ($\textbf{S}_p$) and the final target $\textbf{T}$ in a relative opposite location, as shown in Fig. \ref{Fig2}.  At the center of the circumference a fixed obstacle is placed, which has the approximate size of a pedestrian. 

The experiments were performed with the participation of four volunteers in the Motion Capture Laboratory located at the `Instituto Tecnol\'ogico de Buenos Aires'. 

The volunteers were instructed to walk normally from the starting points to the final target. Each volunteer wore a cap with three markers. Throughout the experiment, each pedestrian had to walk less than 500 m (not continuously)  inside the measurement area at normal speed and with no physical contact. Under these conditions, the experimental protocol did not involve any risks, protecting the integrity, privacy, and confidentiality of the research subjects.

The position of the markers was recorded using
the commercially available technology from Optitrack \textsuperscript{\textregistered}. Each marker position was captured by 16 Flex3\textsuperscript{\textregistered} cameras located throughout the recording environment at 33 frames per second and then processed using the Motive: Body\textsuperscript{\textregistered} software. Pedestrians were tracked using the position of the three markers, but we considered the position of the pedestrian as that of the highest (central) one and the others were used only for reconstruction when, for very short periods, the acquisition system failed to record the position of the central one. The precision of the technology locating a marker in the 3-D space was 1 cm.

The 3-D spatial trajectories obtained was further processed. First, only the two components belonging to the horizontal plane were kept and the height were ignored. Then,  in order to neglect the natural swaying of human walking, a Fourier filter was applied to each of the two horizontal components of the trajectories $\textbf{r}(t)~=~[x(t),~y(t)]$.

\begin{figure*}[ht]
\begin{center}
\centerline{\includegraphics[width=0.6 \textwidth]{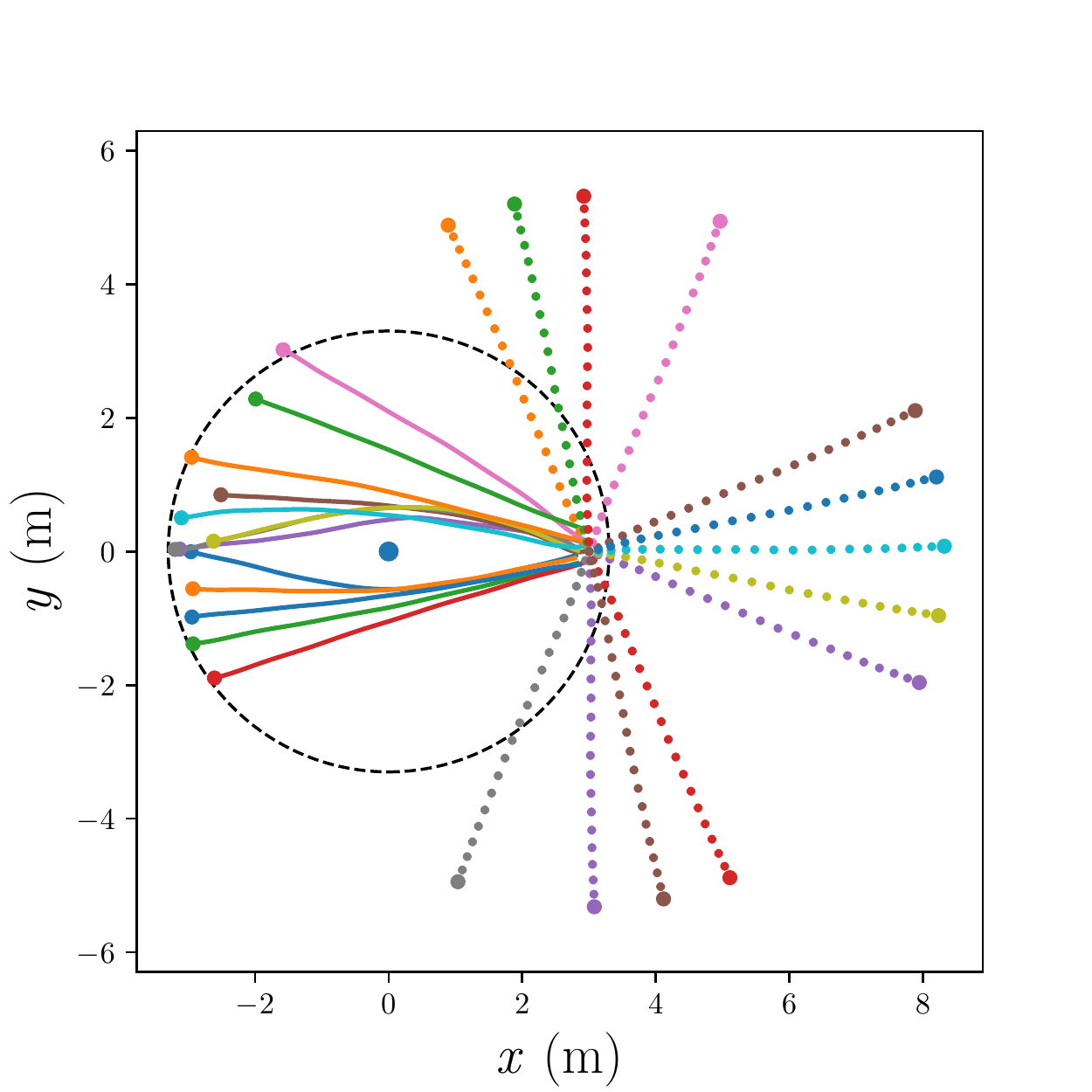}}
\caption{Experimental trajectories (solid lines) and rotated trajectories (dashed lines). The points indicate the initial position of trajectories. The final target for all trajectories  is located at $(x,~y)=$ (0, 3 m) coordinates.}
\label{Fig3new} 
\end{center}
\end{figure*}

Finally, 13 clean trajectories were selected from all the pedestrians, which are displayed as solid lines in Fig.\ref{Fig3new}.

The starting points ($\textbf{S}_p$) near $y \sim 0 $ produced trajectories with detours needed for avoiding the obstacle. However, in the extreme starting locations ($|y| \gtrsim 2 $), the trajectories were almost straight because a direct trajectory toward the final target would not intersect the obstacle, and thus it would not be necessary to dodge it.

In order to have a complete set of trajectories, we replicate these extreme trajectories by rotating them around the final target, spanning the rest of the angular positions, from  which the simulated pedestrian can move directly toward the final target without performing any avoidance. We choose 13 replicated and rotated trajectories in strategic positions that can be seen as dashed lines in Fig. \ref{Fig3new}.

Summarizing, the 26 experimental trajectories in Fig. \ref{Fig3new} contain all the information for a pedestrian approaching a target from any angular position, having or not an obstacle to avoid. Note that this information can be coded in the particle's system of reference (see Sec. \ref{sec2p2}) and thus it is general, in the sense that the relative position of the obstacle can be arbitrary.

\subsection{A neural network instance of $\textbf{H}$}
\label{sec2p3}

We define the input state seen from the particle that will allow us to compute the output action ($\textbf{v}^s$ in eq. \ref{first order equation}) as the temporary target from eq. (\ref{Heuristic}).

Input / output pairs will be obtained from the experimental set of trajectories. Then, a general regression neural network will take these examples for predicting new outputs as from the simulated environment (inputs).

\subsubsection{Input}

In a two-body problem, we can consider the  pedestrian $i$, who has position $\textbf{r}_i(t)$ at time $t$ whose goal is $\textbf{T}_i$, and any other arbitrary pedestrian or obstacle $j$ with position $\textbf{r}_j(t)$. 
We postulate a continuous input state given by the vector $\boldsymbol{\xi}_{ij}(t)$ (eq. \ref{Input}).In the case of two particles the dimension of the input space is 6 (2 for particle $i$ and 4 for the other particle). Of course, the input vector can be generalized; if there were more particles, its dimension would increase by the amount of 4 for each extra particle. 

\begin{equation}
\label{Input}
\boldsymbol{\xi}_{ij}(t)=[~\hat{v}_i,~\hat{\theta}_{ij},~\hat{d}_{iT},~
\hat{v}_{ij},~\hat{\theta}^{v}_{ij},~\hat{d_{ij}}~]
\end{equation}
\vspace{0 pt}

In what follows we describe the variables of the input space. First, in order to make all the variables compatible, because of their different units, and spanning over different ranges, we define them as dimensionless by rescaling to values $\lesssim 2$.

\begin{figure}[H]
  \begin{subfigure}[t]{0.3\textwidth}
    \centering
	\includegraphics[scale=0.5]{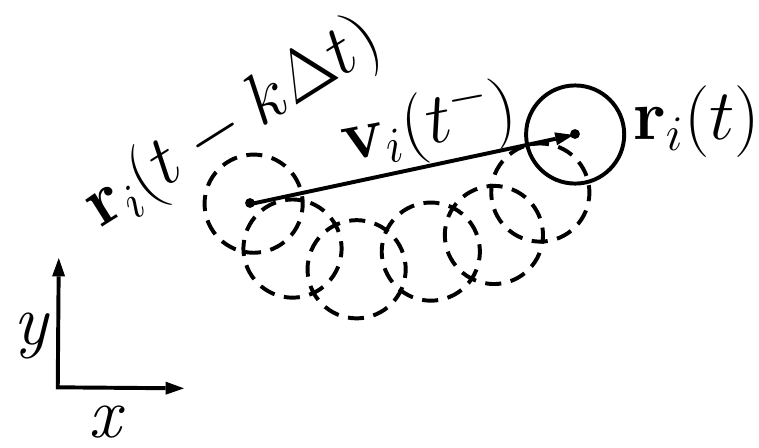}
	\caption{}
  \end{subfigure}
  \hspace{5 pt}
  \begin{subfigure}[t]{0.3\textwidth}
  	\centering
	\includegraphics[scale=0.5]{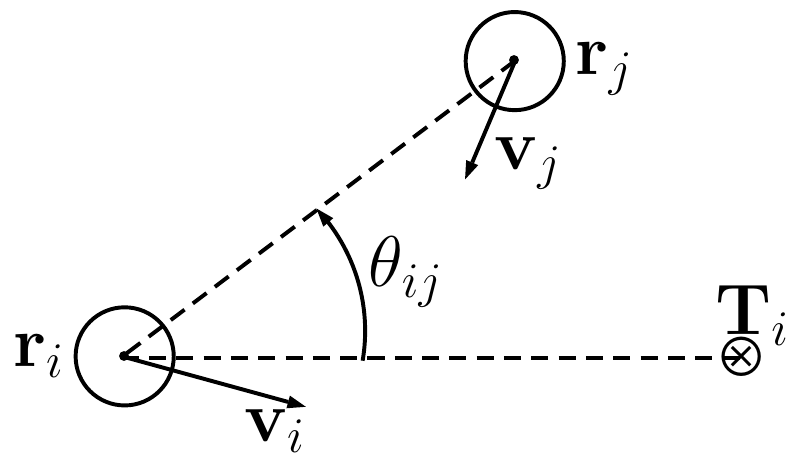}
	\caption{}
  \end{subfigure}
  \hspace{5 pt}
  \begin{subfigure}[t]{1\textwidth}
    \centering
	\includegraphics[scale=.5]{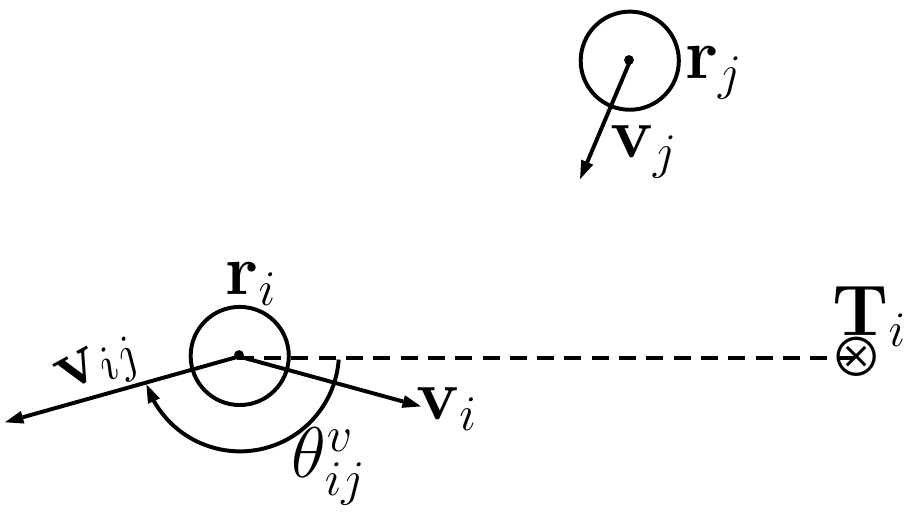}
	\caption{}
  \end{subfigure}
  \caption{Basic quantities needed for defining the input vector ($\boldsymbol{\xi}$).  (a) Past velocity $\textbf{v}_i(t^-)$ of pedestrian $i$. (b) Relative angle $\theta_{ij}$ between both pedestrians. (c) Relative velocity $\textbf{v}_{ij}$ with angle $\theta^v_{ij}$.}
 \label{Fig4new}
\end{figure}

\begin{itemize}
\item $\hat{v}_i$= $\dfrac{|\textbf{v}_i(t^-)|}{ 1.8 \; \text{m/s}}$, where 1.8 m/s is the maximum speed observed in our experiments, $\textbf{v}_i(t^-)$ is the past velocity of pedestrian $i$ at time $t$ calculated as $\textbf{v}_i(t^-) = [\textbf{r}_i(t)-\textbf{r}_i(t-k)]/(k\Delta t)$ (Fig. \ref{Fig4new} (a)).
\vspace{10 pt}

\item $\hat{\theta}_{ij} =
 \begin{cases}
       \ 1 &\quad\text{if} \; \theta_{ij} \geq \pi/2 \\ 
       \ -1 &\quad\text{if} \;\theta_{ij} \leq -\pi/2 \\ 
       \theta_{ij} \: 2/\pi &\quad\text{otherwise} \\ 
     \end{cases}$
     
where $\theta_{ij}$ is the angle defined between the vectors ($\textbf{T}_i-\textbf{r}_i$) and ($\textbf{r}_j-\textbf{r}_i$) as shown in Fig. \ref{Fig4new} (b) and it lies between the interval $[-\pi,\pi]$. However, the input angle $\hat{\theta}_{ij}$ saturates when $|\hat{\theta}_{ij}|\geq \pi/2$, which makes the particle ignore the obstacles behind it. Also note that this variable takes positive and negative values aiming to distinguish between right and left from the particle looking toward the final target.

\vspace{10 pt}

\item $\hat{d}_{iT} =
 \begin{cases}
       \ d_{iT}/4 \text m &\quad\text{if} \; d_T \leq 8 \text m \\        
       \ 2 &\quad\text{otherwise} \\ 
     \end{cases}$
     
$d_{iT}$ being the distance between the particle and its final target ($d_{iT}\:=\:|\textbf{T}_i - \textbf{r}_i|$).  The saturation value (8 m) causes obstacles beyond that distance to be neglected.

\vspace{10 pt}

\item $\hat{v}_{ij} = \dfrac{|\textbf{v}_{ij}|}{1.8\:m/s}$, where $\textbf{v}_{ij}$ is the relative velocity of $j$ seen from particle $i$ ($\textbf{v}_{ij} \: =\: \textbf{v}_j-\textbf{v}_i$).
\vspace{10 pt}

   
\item $\hat{\theta}_{ij}^{v}= \left\{ \begin{array}{lcccccc}
             -2(\theta^{v}_{ij}+\pi)/\pi  & if & \theta^{v}_{ij}  &<&-\pi/2  \\
              -1 & if &  -\pi/2 &<& \theta^{v}_{ij} &<& 0   \\
              +1 & if &  0 &<& \theta^{v}_{ij} &\leq& \pi/2   \\
             -2(\theta^{v}_{ij}-\pi)/\pi  &  if & \theta^{v}_{ij} &>&\pi/2  \\
             
             \end{array}
   \right.$

\vspace{10 pt}

where $\theta^{v}_{ij}$ is the angle between the vectors  ($\textbf{T}_i-\textbf{r}_i$)  and the relative velocity ($\textbf{v}_j-\textbf{v}_i$) as shown in Fig. \ref{Fig4new} (c). This function saturates for values in the range $|\theta^{v}_{ij}| < \pi/2$ because in this case the particle $j$ would be moving away from particle $i$ and as a result,  there cannot be any collision. 

\vspace{10 pt}

\item $\hat{d}_{ij} =
 \begin{cases}
       \ d_{ij}/4 \text m &\quad\text{if} \; d_{ij} \leq 8 \text m \\        
       \ 2 &\quad\text{otherwise} \\ 
     \end{cases}$
     
\vspace{10 pt}     
     
 where $d_{ij}$ is the Euclidean distance between the particles ($|\textbf{r}_i-\textbf{r}_j|)$.
\end{itemize}

\subsubsection{Output}

\begin{figure}[H]
  \begin{subfigure}[t]{0.5\textwidth}
    \centering
	\includegraphics[scale=0.6]{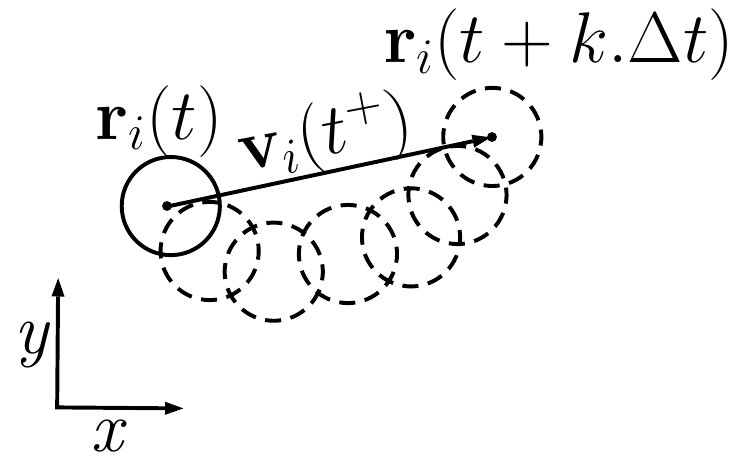}
	\caption{}
  \end{subfigure}
  \hspace{10 pt}
  \begin{subfigure}[t]{0.5\textwidth}
    \centering
	\includegraphics[scale=.6]{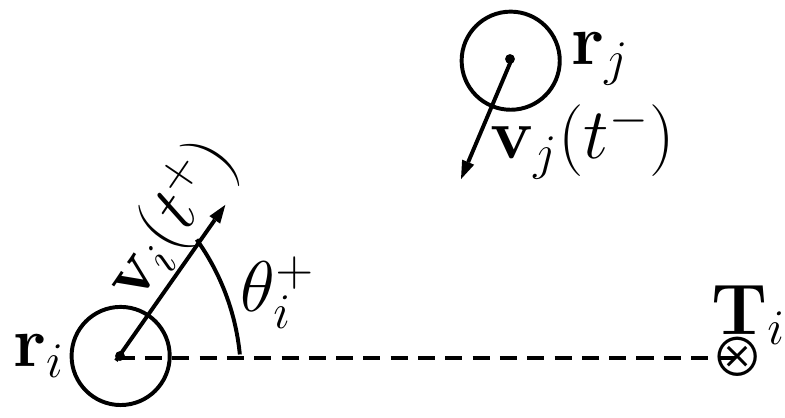}
	\caption{}	
  \end{subfigure}
  \caption{Definition of the output vector ($\boldsymbol{\zeta}$).  (a) Future velocity $\textbf{v}_i(t^+)$ of pedestrian $i$. (b) Angle of the future velocity relative to the direction to the final target ($\theta_i^+$).}
 \label{Fig5new}
\end{figure}

The reaction of the agent will be modeled as its velocity, which in our first-order model will allows us to move it toward its future position (eq. \ref{first order equation}). This velocity will be defined in polar coordinates relative to the direction between the particle and its final target. We call this angle $\theta^+_i$, which is defined between $\textbf{T}_i$ and $\textbf{T}^t_i$ [see Fig. \ref{Fig1} or \ref{Fig5new} (b)]; this definition allows us to have a rotation-invariant data set. In consequence, the output vector has only two dimensions regardless of the number of particles in the system.

\begin{equation}
\label{output}
\boldsymbol{\zeta}_{i}(t) = [\:^1{\zeta}(t),\; ^2{\zeta}(t)\:] = [\:v^+_i,\: \theta_i^+\:].
\end{equation}
\vspace{0 pt}

Here, $v^+_i$ is the speed of particle $i$ for the next time step calculated as the magnitude $v^+_i~=~| \textbf{v}_i|~=~|\textbf{r}_i(t+k \Delta t)-\textbf{r}_i(t)| / (k \Delta t)$ (Fig. \ref{Fig5new} (a)). And $\theta_i^+$ is the angle of the velocity with respect to the direction defined by ($\textbf{T}_i-\textbf{r}_i$) (Fig. \ref{Fig5new} (b)).

If there are no data at time $(t+k \Delta t)$, the output is not calculated and therefore, the corresponding input is not considered.

\subsubsection{The nonparametric neural network} 
\label{GRNN}

The above definitions of input/output pairs (eqs. \ref{Input} and \ref{output} ) can be used for an arbitrary number of interacting particles. However, as stated above, we will consider a simple experimental configuration with one moving particle and one obstacle.

We call $\mathbb{E}=\{\boldsymbol{\xi}(t),\boldsymbol{\zeta}(t)\}$ the experimental set of state/action examples having data points for each time step $t$.

Each one of the two components of the output vector $\boldsymbol{\zeta}(t)$ (eq. \ref{output}) will we approximated by one neural network with output $^{\mu}O \; : \; \mathbb{R}^6\rightarrow\mathbb{R}$, where ${\mu}= {1, 2}$ indicates its polar components, i.e., the speed  ($v_i^+$) and the angle ($\theta_i^+$) respectively.

The neural network we choose is the generalized regression neural network (GRNN) \cite{specht1991general}, which is a type of radial basis function network \cite{park1991universal}. 

The GRNN is a universal interpolator based on nonparametric regression. The  basic idea is that the input of the data samples becomes the center of radial basis neurons in the first layer and their outputs are linearly interpolated in a second layer with weights given by the output of the data samples.

In what follows we explain this concept explicitly for a network with one dimensional output ($O$).
Suppose a training family of ordered pairs $\{\boldsymbol{\xi}_n,\zeta_n\}_{n\leq N}$, then:

\vspace{0 pt}
\begin{equation}
O(\boldsymbol{\xi})=\frac{\sum_{n=1}^N \zeta_n K(\boldsymbol{\xi},\boldsymbol{\xi}_n)}{\sum_{n=1}^NK(\boldsymbol{\xi},\boldsymbol{\xi}_n)}
\end{equation}
\vspace{10 pt}
where
\vspace{0 pt}
\begin{itemize}
\item $O(\boldsymbol{\xi})$ is the prediction value of an arbitrary input vector $\boldsymbol{\xi}$.
\vspace{10 pt}

\item $\zeta_n$ is the activation weight for the pattern layer at neuron $n$.
\vspace{10 pt}

\item $K(\boldsymbol{\xi},\boldsymbol{\xi}_n)=e^{{-l_n}/{2\sigma^2}}$ is the radial basis function kernel.
\vspace{10 pt}

\item $l_n=(\boldsymbol{\xi}-\boldsymbol{\xi}_n)^T(\boldsymbol{\xi}-\boldsymbol{\xi}_n)$ is the square distance between data examples $\boldsymbol{\xi}_n$ and the input vector $\boldsymbol{\xi}$.
\end{itemize}

The only degree of freedom in this neural network is the so-called spread ( $\sigma$), which can be taken as a scalar value for all examples and variables of the input vector.

\section{Simulations}
In this section we describe how the spread ($\sigma$) of both GRNN's was calibrated and we present results showing that with the proposed approach we can to simulate several configurations of a pedestrian avoiding an obstacle, within and beyond the boundaries of the experimental data.
\newline
\subsection{Simulation scheme}
\label{SimScheme}
\begin{figure}[H]  
    \centering
	\includegraphics[scale=0.33]{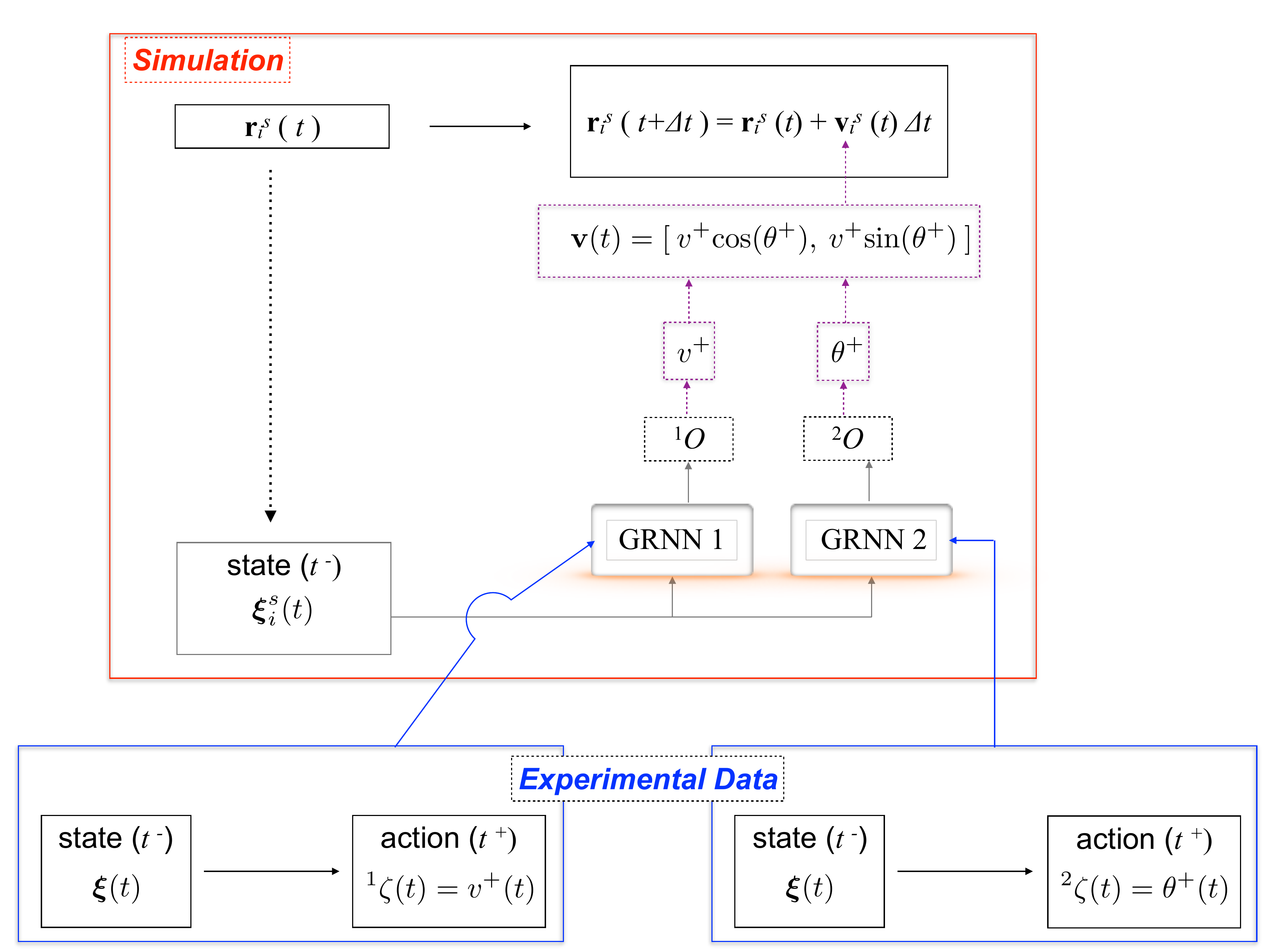}
  \caption{Flow diagram of the simulation procedure by estimating the future velocity of the simulated particle by means of two GRNN's that use data extracted from the same experimental set of trajectories. All symbols were defined in previous sections. Superscript $s$ stands for {\it simulated} positions and velocities.}
 \label{FigXnew}
\end{figure}

The proposed data-driven simulation method is shown in schematic form in Fig. \ref{FigXnew}.

At each time step of the simulation, both GRNN's will provide the speed and angle of the velocity for computing the future position of the particle. In order to do this, the GRNN's will use the experimental data set $\mathbb{E}=\{\boldsymbol{\xi}(t),\boldsymbol{\zeta}(t)\}$, which is provided in the supplementary material. As stated in Sec. \ref{GRNN}, there is only one free parameter for each GRNN: the spread ($\sigma$). In the next section we specify how this parameter was determined.

\subsection{Calibrating the GRNN}

In order to synchronize the simulation with the experimental data, we consider a time step $\Delta t\:=\:1/33\:$s corresponding to the maximum time resolution of the acquisition system ($k=1$ in Fig. \ref{Fig4new} (a) and \ref{Fig5new} (a)). Considering that the training data set is composed of 26 trajectories (shown in Fig. \ref{Fig3new}) of about 97 points each, the number $n$ of input/output patterns is $n \sim 2460$. 

We can write the set of all data points as a collection of 26 subsets corresponding to each trajectory $^{26}\boldsymbol{\xi}_n=\{\textbf{r}_1,\textbf{r}_2,...,\textbf{r}_{26}\}$, where $\textbf{r}_i=[\textbf{r}_i(t)]_{t \leq t_i^f}$ is the succession of pedestrian positions at discrete time steps $\Delta t$ and $t_i^f$ is the final time of $\textbf{r}_i$. The first 13 trajectories correspond to original data, and trajectories from 14 to 26 are replications of the extreme trajectories rotated as explained in \ref{sec2p2}.

Now, for the determination of the spread in each GRNN, we only consider the 13 original trajectories $^{13}\boldsymbol{\xi}_n=\{\textbf{r}_1,\textbf{r}_2,...,\textbf{r}_{13}\}$ and proceed with a leave-one-out cross-validation. We take out each trajectory $\textbf{r}_{i}$ from the set of patterns ($^{13}\boldsymbol{\xi}_n$) to reconstruct the same trajectory ($\textbf{r}^s_{i}$) by simulating it with the methodology described in Fig. \ref{FigXnew} considering the remaining trajectories as data examples for the GRNN's. Because each simulated trajectory needs two neural networks (one for each polar coordinate) the global error will be a function of both of them.  We call $E_i(\sigma_1,\sigma_2)$ the error when comparing $\textbf{r}_i$ with $\textbf{r}^s_i$. Then, $E(\sigma_1,\sigma_2)=\frac{\sum_{i=1}^{13}E_i(\sigma_1,\sigma_2)}{13}$ is the global error for these spread values. 

As boundary conditions for each simulated trajectory $\textbf{r}^s_i$ we take the initial position ($\textbf{r}^s_i(0)=\textbf{r}_i(0)$) and velocity  $(\textbf{v}^s_i(0)=[\textbf{r}_i(\Delta t)-\textbf{r}_i(0)]/\Delta t$ equal to the experimental ones. The final target is selected as the last position of the experimental trajectory [$\textbf{T}_i = \textbf{r}_i(t_i^f)$].

We define two different average error functions between simulated and experimental trajectories, one based on the minimum distance to the obstacle, located in $\textbf r_{obs} = (0,0)$, which we call $E_d(\sigma_1,\sigma_2)$ defined by eq. (\ref{eq6}), and the other is the mean of the difference of position at the same time step and we call it $E_t(\sigma_1,\sigma_2)$ defined by eq. (\ref{eq7}).

\begin{equation}\label{eq6}
E_d(\sigma_1,\sigma_2)=\frac{1}{13}    \sum_{i=1}^{13}  |\text{min}_{t}( | \textbf{r}_i^s(t) - \textbf{r}_{obs} | ) - \text{min}_{t} ( | \textbf{r}_i(t)- \textbf{r}_{obs} | ) |,
\end{equation}

\begin{equation}\label{eq7}
E_t(\sigma_1,\sigma_2)=\frac{1}{13}    \sum_{i=1}^{13}    \sum_{t\leq t_i^f}\frac{{ | \textbf{r}_i^s(t)-\textbf{r}_i(t) |}}{ t_i^f }.
\end{equation}
\vspace{0 pt}

First, we consider the simplified case in which the same spread value is used for both neural networks ($\sigma_1=\sigma_2 \equiv \sigma$). This assumption can be made because the input space is the same for both networks and the spread is a measure of how many patterns are considered for estimating the output.

\begin{figure}[h]
  \begin{subfigure}[b]{0.5\textwidth}
    \includegraphics[scale=0.33]{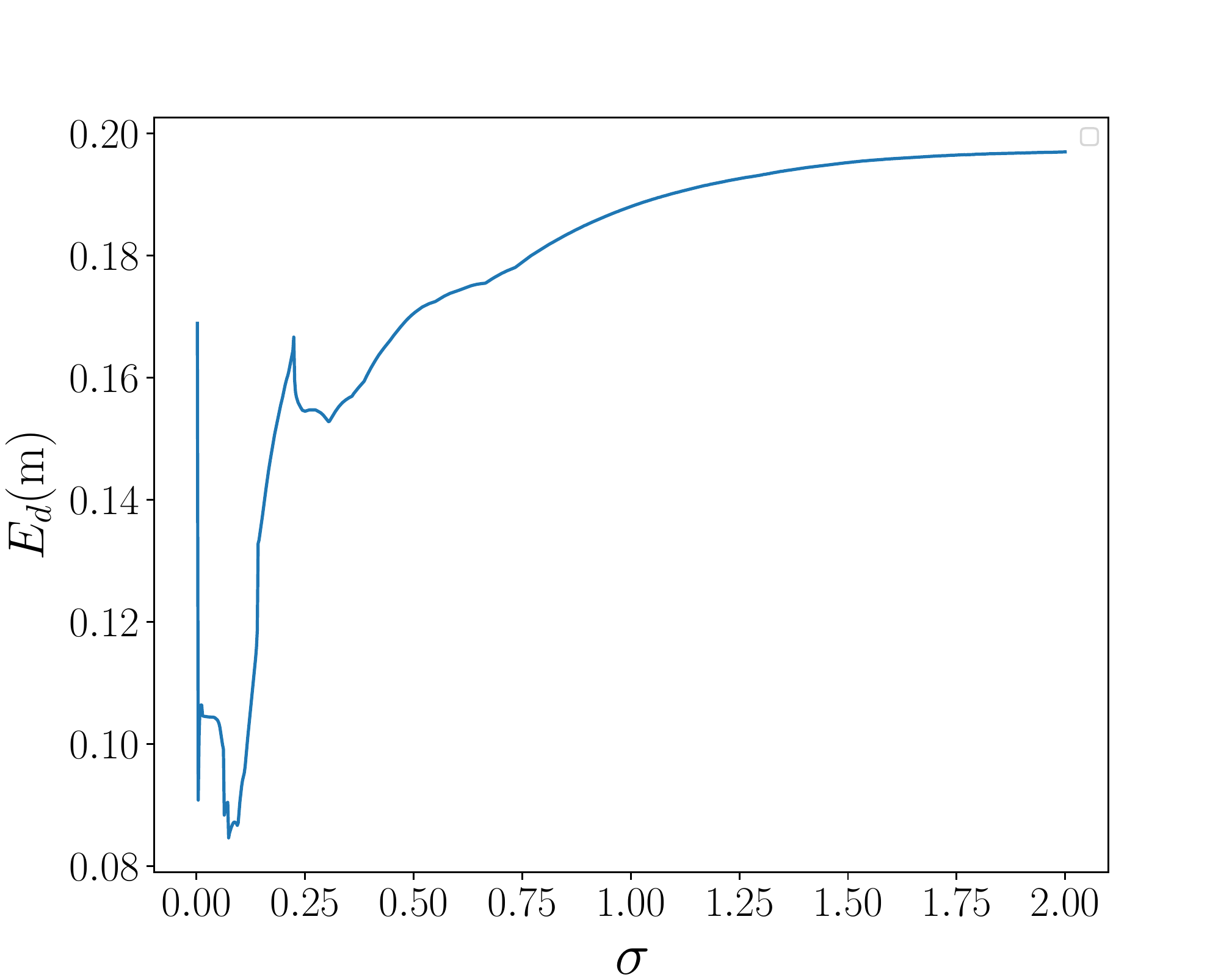}
    \caption{}
  \end{subfigure}
  \begin{subfigure}[b]{0.5\textwidth}
    \includegraphics[scale=0.33]{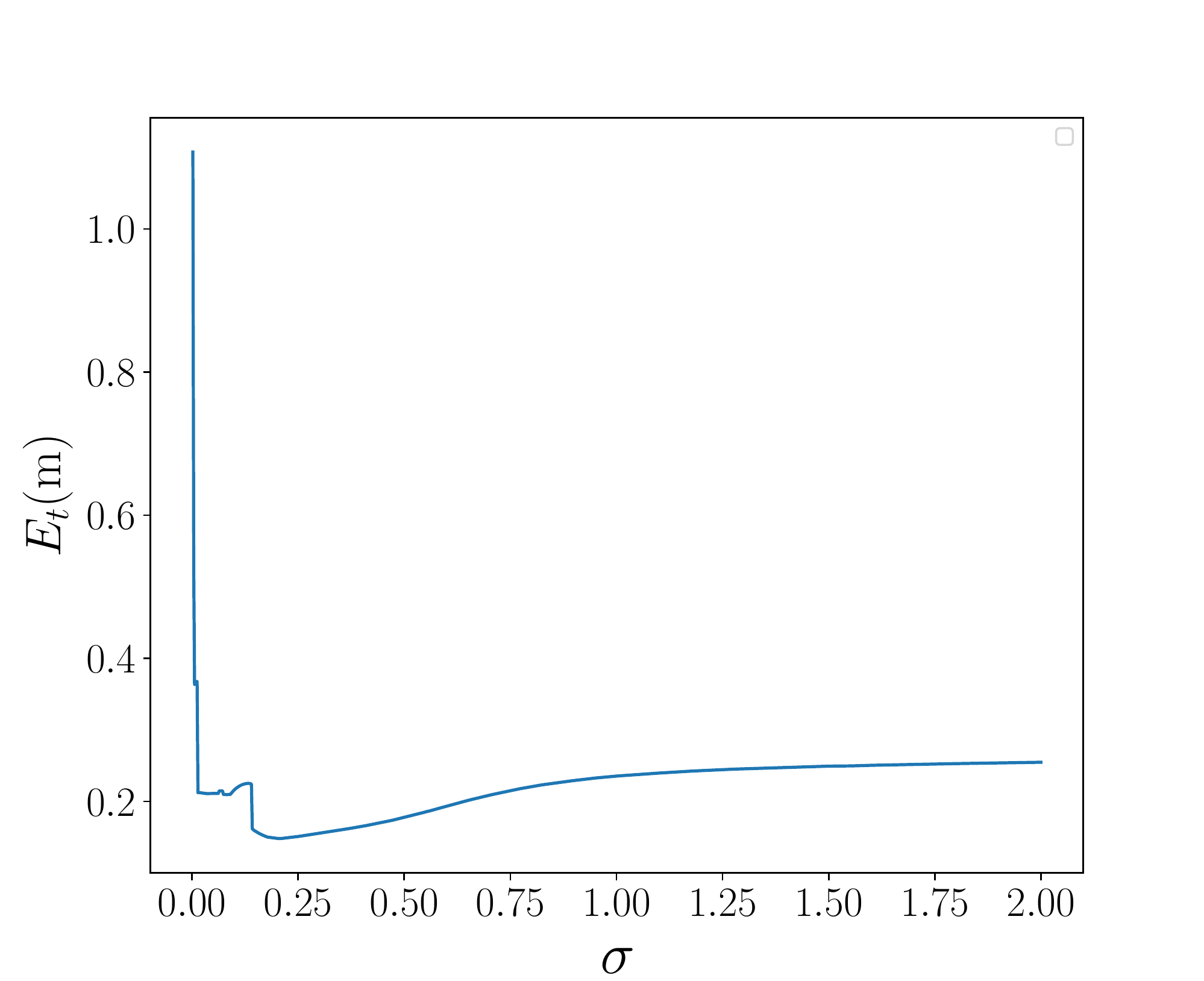}
    \caption{}
  \end{subfigure}
   \caption{Measures of the error for the simulated trajectories comparing with the experimental ones as a function of the GRNN parameter $\sigma$. (a) Minimum distance to obstacle error: $E_d$ (eq. \ref{eq6}). (b) Average microscopic difference between trajectories: $E_t$ (eq. \ref{eq7}).}
   \label{Fig7new}
\end{figure}

Figure \ref{Fig7new} shows the results. The optimum spreads found were: $\sigma = 0.074$ for the distance-to-obstacle error $E_d=0.08$ m, and $\sigma = 0.20$ for the frame-to-frame error $E_t=0.15$ m. 

These spread values provide us with a range of usable values in the proposed network. Interestingly, it contains $\sigma~=~<d_{fn}>~ = ~0.11 $, which is the mean distance between first neighbors (excluding points from the same trajectory) in the input space. Then, we can state that the GRNN's having better performance on the data are those that take into account the closest data points in the input space, with respect to the new input to be predicted.  

\begin{figure}[H]
 \centering
    \includegraphics[scale=0.5]{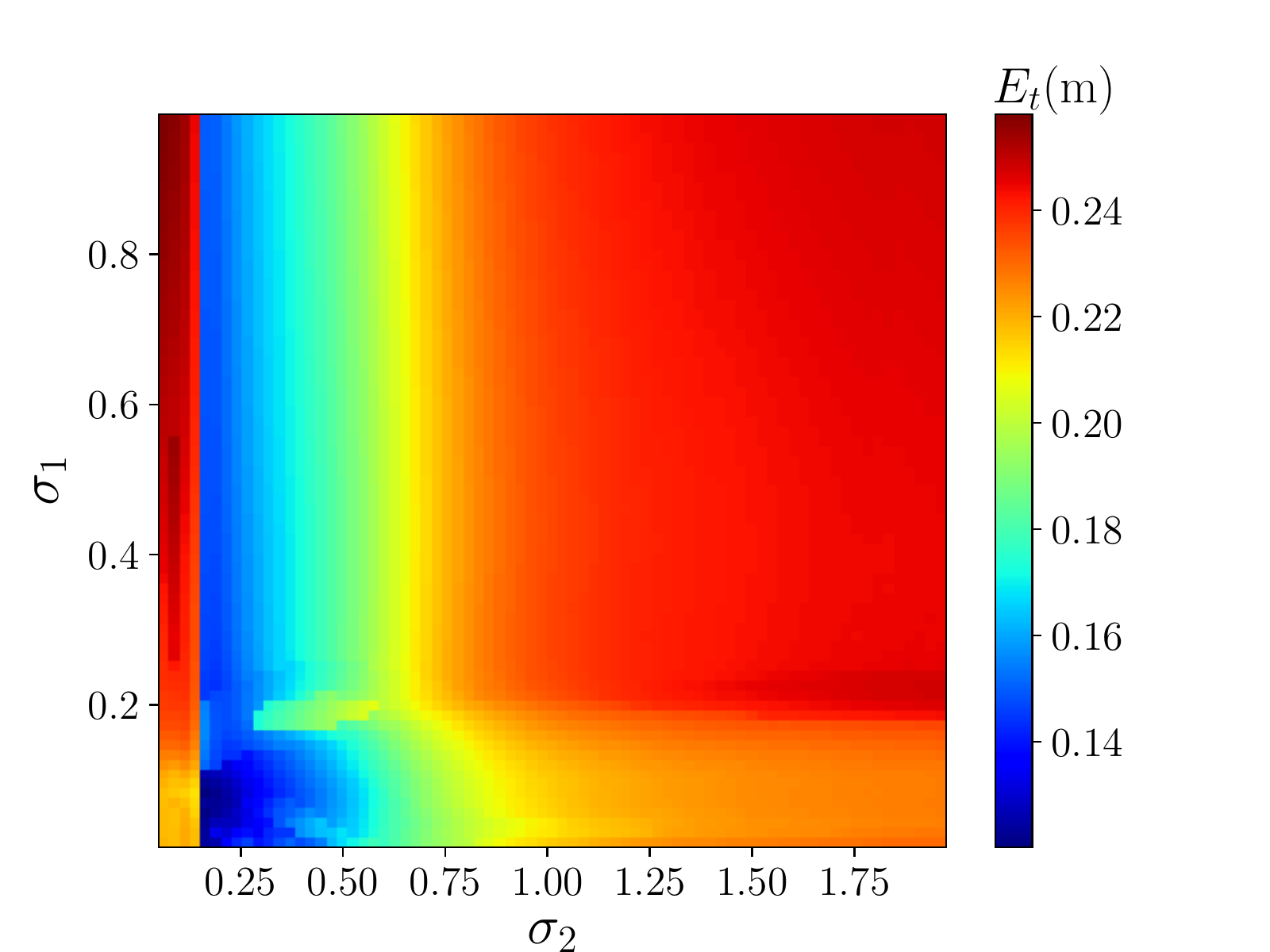}
    \caption{Average microscopic error: $E_t$ (eq. \ref{eq7}) as a function of the spreads of both GRNN's.}
    \label{Fig8new}
  \end{figure}

Now, in order to improve the $E_t$ error obtained above we need a better approximation of the speed, consequently, we relax the $\sigma_1=\sigma_2$ constraint and explore this error as a function of the two variables. The heat map plot in Fig. \ref{Fig8new} displays the minimum value of the error $E_t=0.12$ m at $\sigma_1=0.16$ and $\sigma_2=0.08$. We can see that the decoupling of both GRNN's leads to a better approximation of the data considering a microscopic comparison between simulated and experimental trajectories. Again, the spread values obtained are comparable with the mean distance between first neighbors in the input space.

The simulated trajectories obtained with these parameters are presented in the next section.

\subsection{Results} 
With the spreads found in the previous section and the trajectory data described in Sec. \ref{sec2p2} and \ref{sec2p3}, we analyze the performance of the simulations using the proposed method described in Sec. \ref{SimScheme}.
 
\begin{figure}[H]
  \begin{subfigure}[b]{0.5\textwidth}
   \includegraphics[scale=0.5]{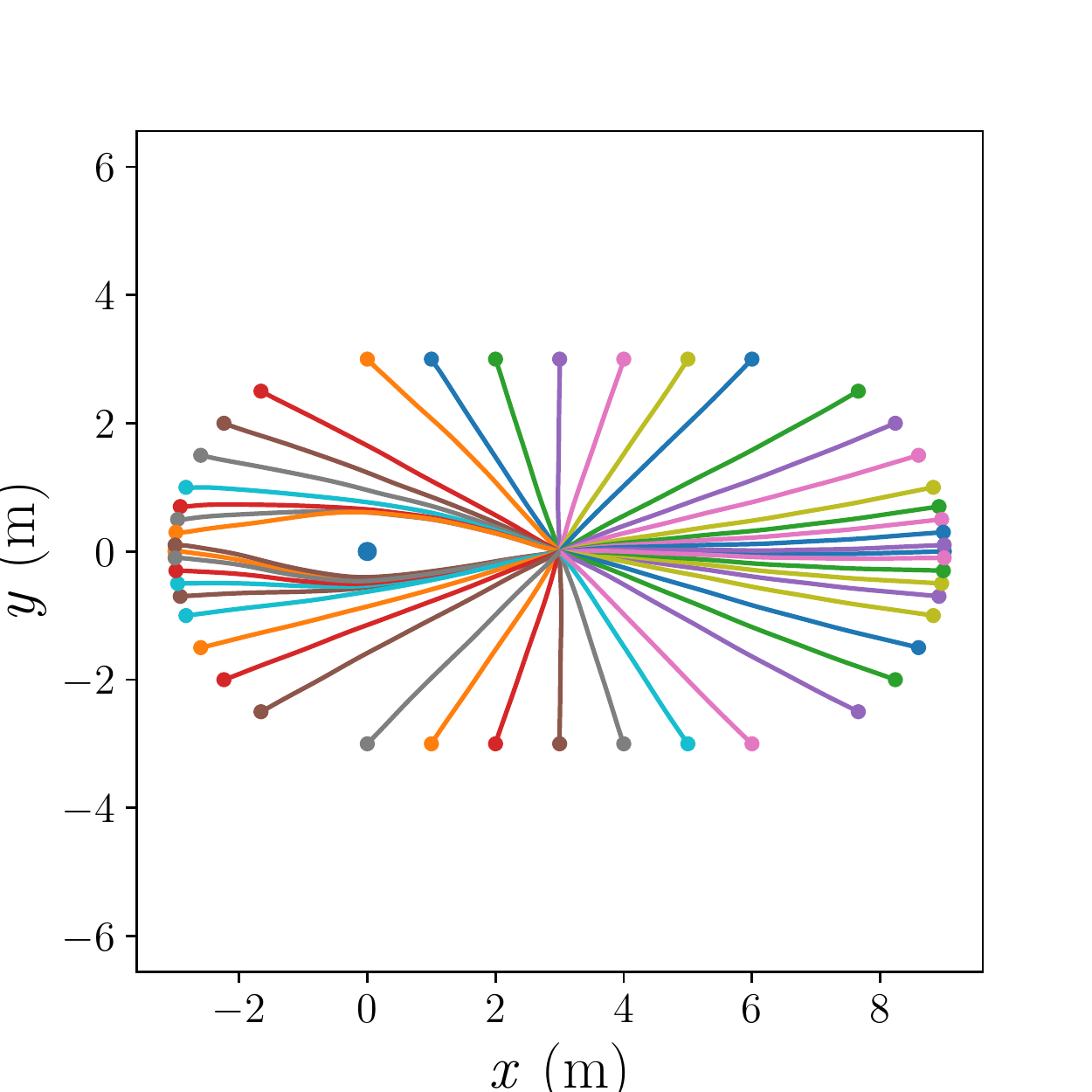}
   \caption{}
  \end{subfigure}
  \begin{subfigure}[b]{0.5\textwidth}
    \includegraphics[scale=0.5]{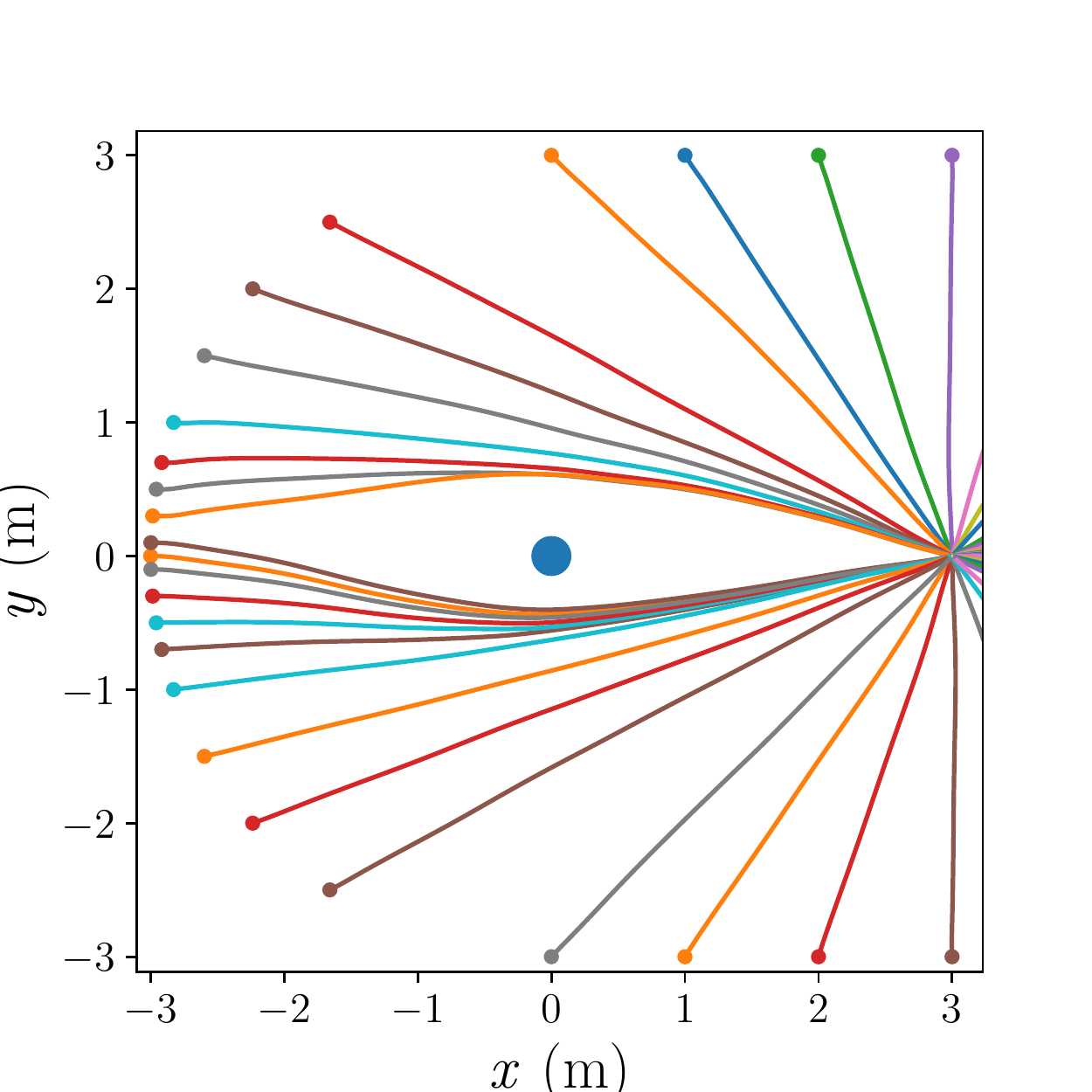}
    \caption{}
  \end{subfigure}
  \caption{Simulated trajectories with $\sigma_1=\sigma_2=\sigma=0.11$. (a) Complete view of the simulated system. (b) Zoom over the avoidance region.}
   \label{Fig9new} 
\end{figure}

The system to be simulated consists of a fixed obstacle located in $(x,y)=(0,0)$ and a final target in $\textbf{T}_i=(3~\text{m},0)$ for all $i$. Forty-eight new particles were simulated once at a time, with initial positions at 6 m from $\textbf{T}_i$ as shown in Fig. \ref{Fig9new}. Neither of these trajectories is equal to the experimental ones. In all cases, the initial velocity points toward $\textbf{T}_i$ and has an initial speed of $v_i=1.27$ m/s, which is the arithmetic mean of the initial speeds of all the experimental trajectories.

First, we present the results corresponding to the case in which both spreads are considered equal to $\sigma=0.11$.  In Fig. \ref{Fig9new} the smoothness and continuity of the trajectories with respect to the initial positions can be seen. Also note that the minimum distances of the trajectories from the obstacles are similar to those from our experiments ($<d_{min}> ~ \sim ~ 0.5m$) and from other papers \cite{jia2019experimental, parisi2016experimental}.

It should be noted that only potentially colliding trajectories produce a detour for avoiding the obstacle, while the rest of the particles describe straight trajectories toward the target. Also, if the obstacle were located in another position (at similar distance from the target), the trajectory patterns would rotate accordingly, because the input state and output action are defined in a coordinate system relative to the particle, i.e., polar coordinates taking the zero angle axes as the direction from the particle toward the target ($\mathbf{T}_i - \mathbf{r}_i$). Thus, the results do not depend on the absolute position of the obstacle.

\begin{figure}[H]
  \begin{subfigure}[b]{0.5\textwidth}
   \includegraphics[scale=0.5]{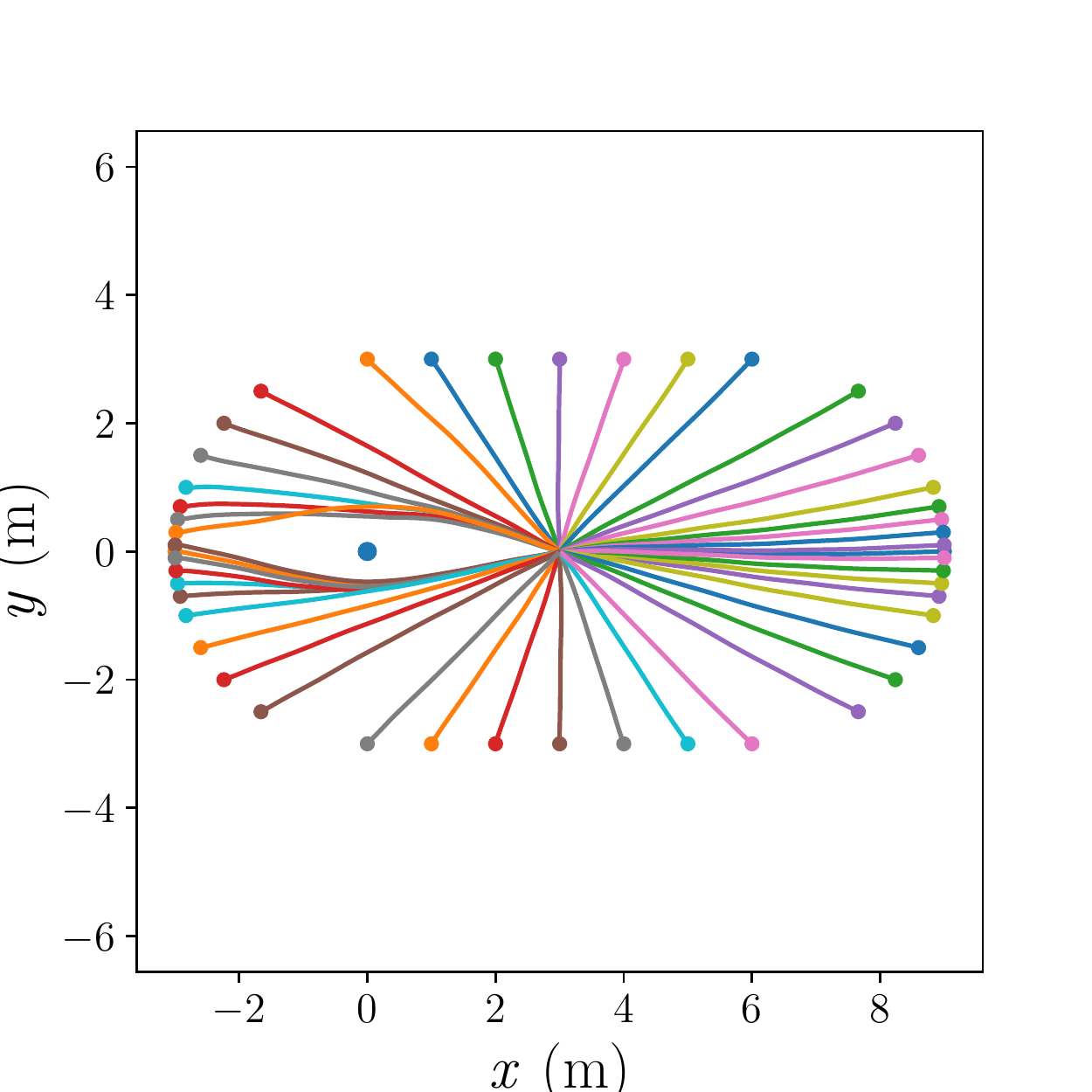}
   \caption{} 
  \end{subfigure}
  \begin{subfigure}[b]{0.5\textwidth}
    \includegraphics[scale=0.5]{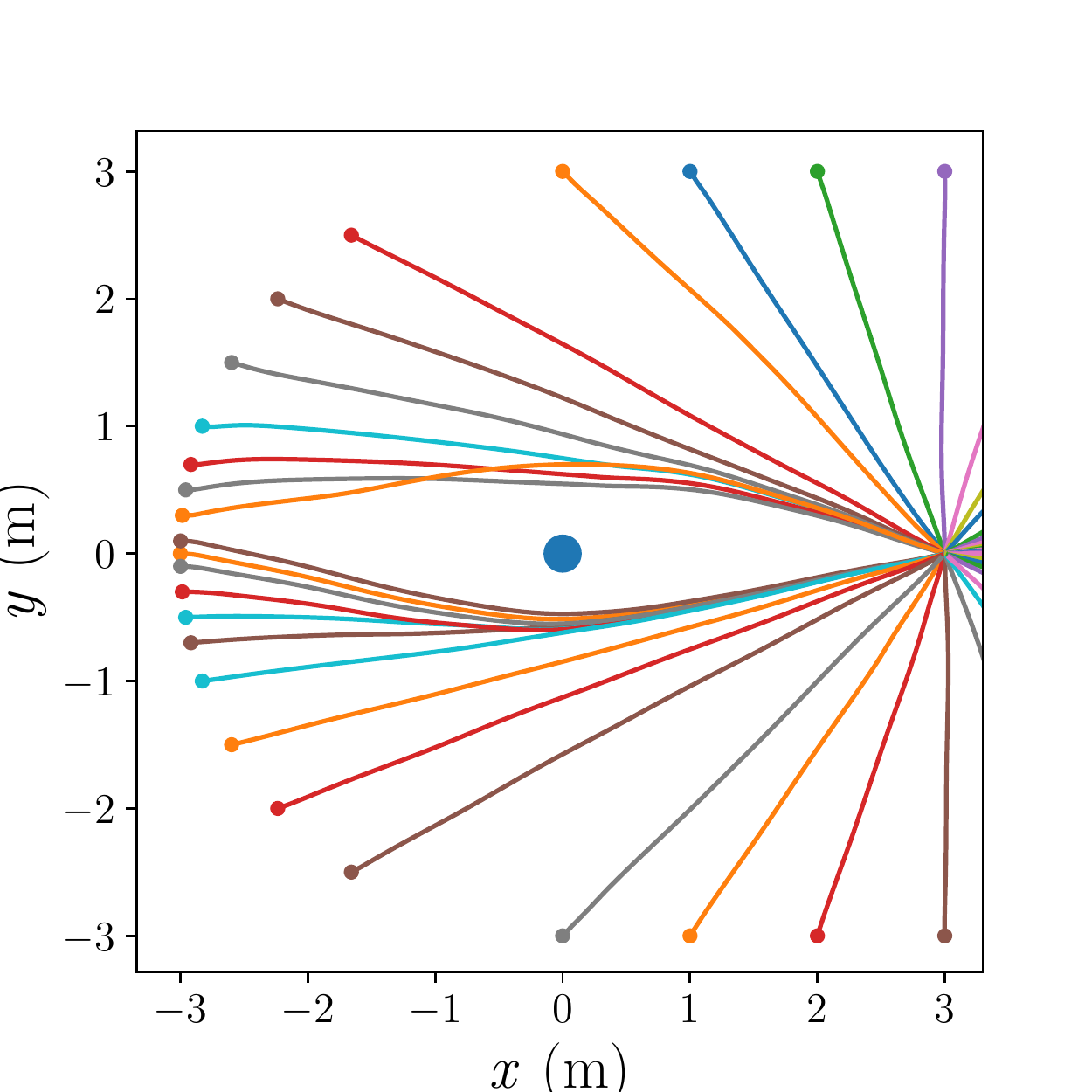}
    \caption{} 
  \end{subfigure}
  \caption{Simulated trajectories with $\sigma_1=0.16$ and $\sigma_2=0.08$. (a) Complete view. (b) Zoom over the avoidance zone.}
  \label{Fig10New}
\end{figure}

Second, we explore the same configuration, also simulating one particle at a time, but using the different spread values for each GRNN found in the previous section ($\sigma_1=0.16$ and $\sigma_2=0.08$). The results are presented in Fig.\ref{Fig10New}. 

Also in this case, the simulated trajectories correctly describe the avoidance behavior. However, some differences can be observed; for example, there is one trajectory that slightly crosses over other neighbors' trajectories.  This crossing is also observed in the experiments as is shown in Fig. \ref{Fig3new}. 

An important consequence is that different values of $\sigma$ can lead to different avoidance behaviors, which can be used for simulating a heterogeneous population of virtual pedestrians. Of course, another way of doing this would be to use a different data set of trajectories for the GRNN of each simulated agent.

\subsection{Extrapolation to more complex obstacles}

In order to see whether our model has any prediction capacity, we stress the proposed methodology by simulating some configurations different than those of the experimental setup. In particular, we consider more complex obstacles composed of several simple ones.

For these simulations we choose the variant of using the same $\sigma=0.11$ for both GRNN's.

\begin{figure}[H]
  \begin{subfigure}[b]{0.5\textwidth}
   \includegraphics[scale=0.5]{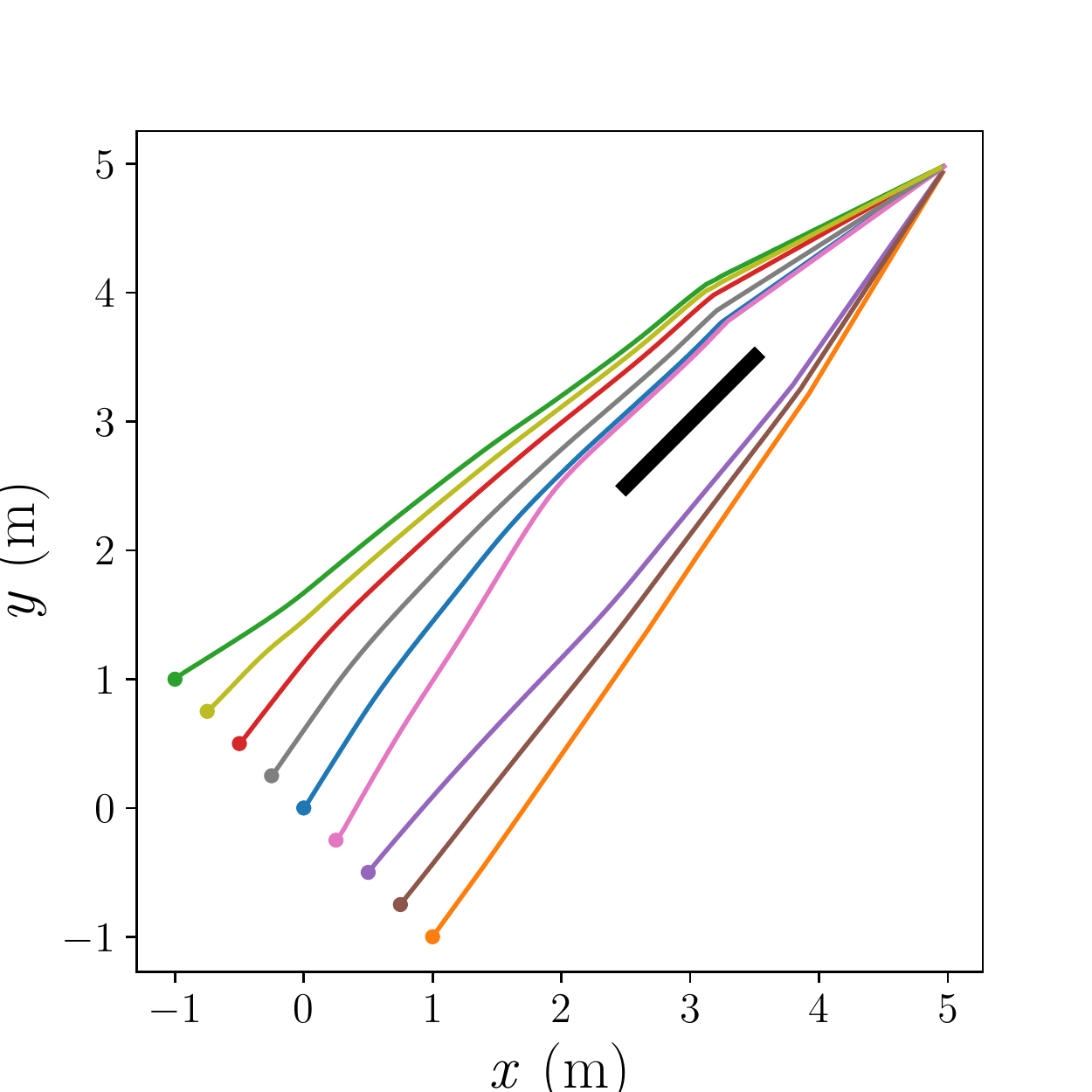}
   \caption{} 
  \end{subfigure}
  \begin{subfigure}[b]{0.5\textwidth}
    \includegraphics[scale=0.5]{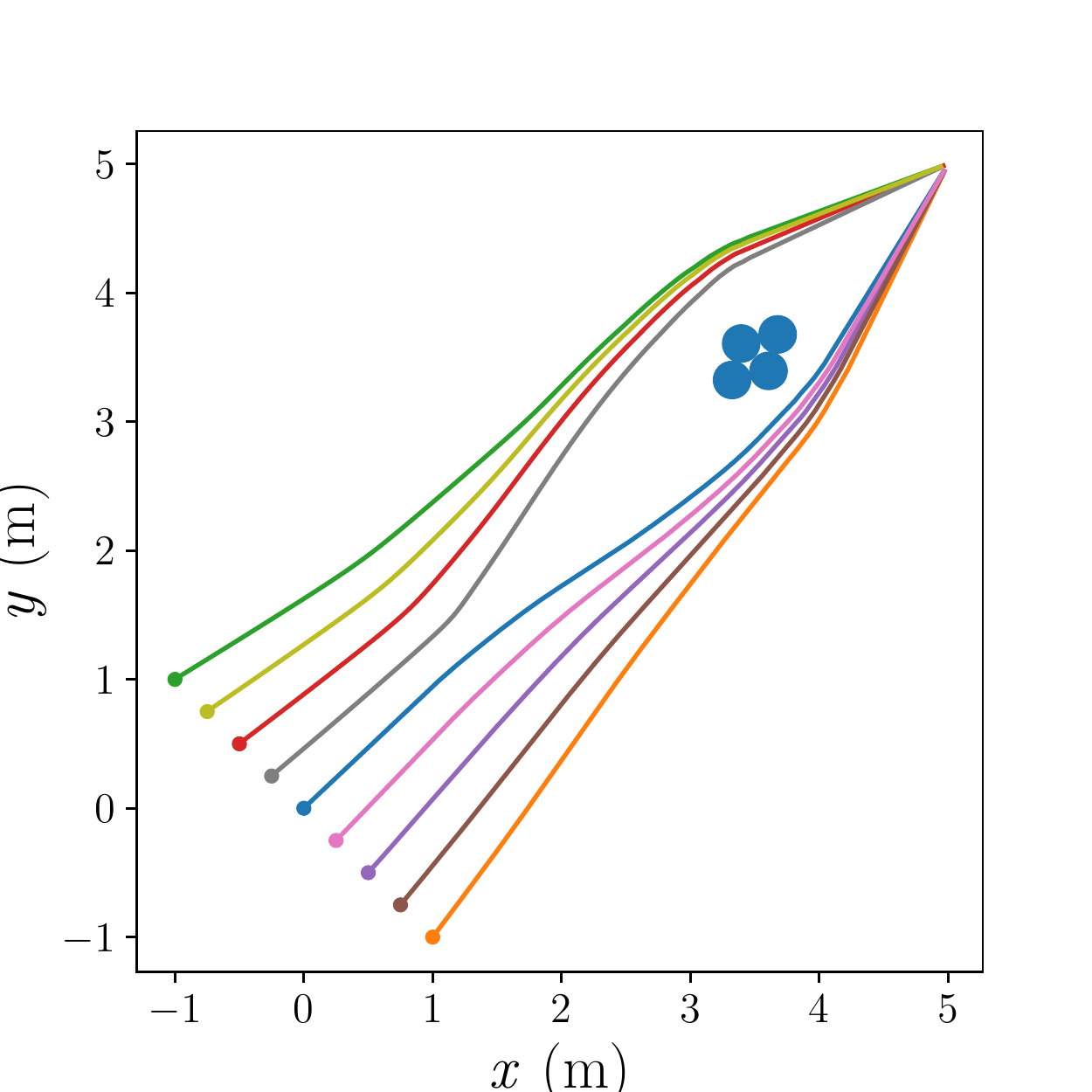}
    \caption{} 
  \end{subfigure}
  \caption{Simulated trajectories in previously unseen scenarios. (a) A wall-like obstacle. (b) A larger obstacle built from four basic obstacles }
  \label{Fig11New}
\end{figure}

In Fig. \ref{Fig11New} (a) a wall-like obstacle of length 1.4 m is avoided by the simulated particles. The wall is oriented along the direction of trajectories. And the simulated pedestrian considers the closest point over the wall in its field of view as the obstacle to be avoided.

Another configuration of a larger obstacle composed of four basic obstacles is presented in Fig. \ref{Fig11New} (b). In this case, the simulated agent considers the closest obstacle and reacts in consequence, following the proposed method. Also here, it can be observed that the particles dodge the obstacle at reasonable distance.

In both cases,  it can be seen that the simulated particles describe natural trajectories when avoiding larger obstacles. It is also demonstrated that arbitrary rotation of the system does not affect the performance of the method.

\section{Conclusions and perspective}

The collision avoidance problem considering one moving pedestrian and one obstacle was studied with the technology from a motion capture laboratory allowing high-precision tracking of trajectories.

A data-driven model that uses a generalized regression neural network (GRNN) was proposed as a general method for simulating pedestrian dynamics. As a first approach, it was implemented with a simple configuration studied experimentally, consisting of one pedestrian avoiding a fixed obstacle.

The advantage of the GRNN is that it only has one free parameter and no training phase is needed, because the input/output patterns are used directly by this neural network. 

The proposed simulation scheme allows us to reproduce the experimental data and generalize to other scenarios not explicitly contained in these data used to feed the GRNN. In this sense, the methodology proposed is invariant under rotations of the relative particle - obstacle positions. Thus we can claim that the data-driven simulation of the general problem of avoiding one narrow obstacle at large and medium distances has been solved.

We provide the model along with a range of spread values, the free parameter of the GRNN.

Our model is ready for considering more particles. Thus, in future work we will present results of navigation in more populated environments at medium and also at high densities, where contact and competitiveness could be present.

\section*{Acknowledgements}
The authors acknowledge financial support via project PID2015-003 (Agencia Nacional de Promoci\'on Cient\'ifica y Tecnol\'ogica, Argentina; Instituto Tecnol\'ogico de Buenos Aires; Urbix Technologies S.A.) and from ITBACyT-2018- 42 (Instituto Tecnol\'ogico de Buenos Aires). Also, we are grateful to Marcela Guerrero for her invaluable collaboration setting-up the Motion Capture system and to the four volunteers of the experiments.

\section*{References}

\bibliography{biblio}

\end{document}